# A NEW LINEAR INDUCTIVE VOLTAGE ADDER DRIVER FOR THE SATURN ACCELERATOR

M. G. Mazarakis, R. B. Spielman, K. W. Struve, F. W. Long
Sandia National Laboratories, Albuquerque, NM 87185, USA

*Abstract*

Saturn is a dual-purpose accelerator. It can be operated as a large-area flash x-ray source for simulation testing or as a Z-pinch driver especially for K-line x-ray production. In the first mode, the accelerator is fitted with three concentric-ring 2-MV electron diodes, while in the Z-pinch mode the current of all the modules is combined via a post-hole convolute arrangement and driven through a cylindrical array of very fine wires. We present here a point design for a new Saturn class driver based on a number of linear inductive voltage adders connected in parallel. A technology recently implemented at the Institute of High Current Electronics in Tomsk (Russia) is being utilized[1].

In the present design we eliminate Marx generators and pulse-forming networks. Each inductive voltage adder cavity is directly fed by a number of fast 100-kV small-size capacitors arranged in a circular array around each accelerating gap. The number of capacitors connected in parallel to each cavity defines the total maximum current. By selecting low inductance switches, voltage pulses as short as 30-50-ns FWHM can be directly achieved.

The voltage of each stage is low (100-200 kV). Many stages are required to achieve multi-megavolt accelerator output. However, since the length of each stage is very short (4-10 cm), accelerating gradients of higher than 1 MV/m can easily be obtained. The proposed new driver will be capable of delivering pulses of 15-MA, 36-TW, 1.2-MJ to the diode load, with a peak voltage of ~2.2 MV and FWHM of 40-ns. And although its performance will exceed the presently utilized driver, its size and cost could be much smaller (~1/3). In addition, no liquid dielectrics like oil or deionized water will be required. Even elimination of ferromagnetic material (by using air-core cavities) is a possibility.

## 1 INTRODUCTION

Saturn is a pulsed power accelerator presently in operation at Sandia[2]. It represents a modification of the old PBFA I[3] accelerator used for ion fusion research. It is named Saturn because of its unique multiple ring diode design which is utilized while operating as an x-ray bremsstrahlung source. As an x-ray source Saturn can deliver to the three-ring e-beam diode a maximum energy of 750 kJ with a peak power of 32 TW providing an x-ray exposure capability of 5 x $10^{12}$ rads/sec over a 500cm$^2$ area. As such it is the highest power electrical driver for bremsstrahlung production in the world. As a z-pinch driver Saturn can deliver up to 8 MA to a wire or gas-puff load. A 700-kJ total x-ray output was obtained from aluminum or tungsten wire pinches, and a 60-70-kJ K-line radiation from aluminum

## 2 PRESENT ACCELERATOR CONFIGURATION

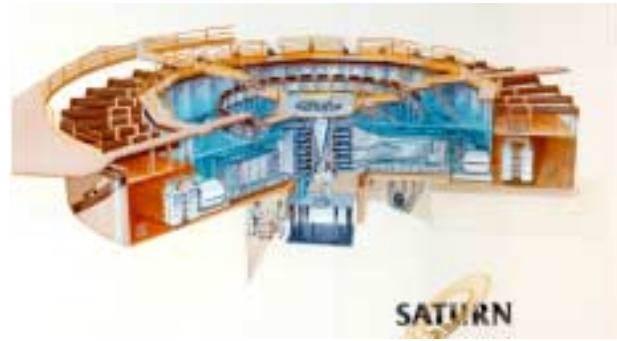

Fig. 1 Schematic representation of the present Saturn configuration

Figure 1 is a schematic representation of the Saturn configuration presently in effect. It utilizes conventional pulse power architecture. The driver starts with 36 Marx generators as the principal energy source. Then ensues a cascade of pulse compression stages to convert the microsecond FWHM Marx output to the 50-ns final pulse that powers the electron diode or the z-pinch load. The power flow follows the following stages: The microsecond pulses are stored in 36 water intermediate store capacitors. From there through 36 triggered gas switches they charge 36 pulse-forming lines which through 36 self-breaking water switches launch the short pulses into 36 triplate water transmission lines and impedance matching transformers. Finally, the short (now ~50 ns) pulse is applied to the diode through a water-vacuum insulating ring interface. The 36 vertical triplate transmission lines are connected to three horizontal triplate disks in a water convolute section. In the vacuum section the power is split into three parts and feeds three conical-triplate magnetically insulated transmission lines (MITL) that power the three separate rings of the e-diode. A final pulse compression of 25:1 is achieved. To accomplish all that, the device requires substantial dimensions. The overall diameter is 30 m and the height 5 m. The 36 Marxes are immersed in a 250,000-gallon oil-filled annular tank while the rest of the device is in 250,000 gallons of deionized water. In our proposed design, utilizing the fast-pulser technology[5], the entire Saturn device will be smaller than 17 m in diameter and will be only 5 m in height. In addition, no water and oil tanks are required. Except for the 5-m high, 4-m diameter cylindrical central section which is in vacuum, the entire accelerator is in the air and readily serviceable.

The three diodes are concentric and, hence, have different power requirements. The innermost diode requires less power than the outside one to achieve the same x-ray area average output. The diode design is such that a power partition with a ratio of 3:2:1 from outside ring to center ring is required. The bottom triplate MITL feeds the largest outer diode ring while the top MITL is connected with the innermost diode ring.

## 3 NEW SATURN DESIGN BASED ON OUR FAST PULSER ARCHITECTURE

In our new design we utilize the revolutionary technology of the fast pulser[4]. We select 24 modules of ~700 kA, 2.2 MV each to produce a total current of the order of 14 MA. The novelty of the design is that each of the 24 modules is a self-magnetically-insulated voltage adder. No liquid or solid insulator is utilized between the inner cathode electrode and the outer anode cylinder. A coaxial geometry is adopted. The stages of each module voltage adder are of relatively low voltage (100 kV). Therefore, in principle a 20-stage structure would be enough to obtain a 2 MV output. However, because we adopted a design where the output impedance is equal to the characteristic impedance of the pulser[4], the peak voltage of each stage will not exceed 55 kV. Hence, 40 stages per module will be necessary. The prime power per stage is a circular array of 24, 30-kA, maximum-current fast capacitors. The inductance and capacitance of each capacitor is 25 nH and 11 nF respectively, which provide a very fast, $\sqrt{LC} = 17$ ns, time constant for the system. Hence, no further pulse power compression is required. The power-pulse FWHM from its onset has the required width to be applied directly to the diode.

The capacitor arrays are switched to the accelerating gaps through very low inductance, ~ 1 nH, externally triggered switches. Two candidates are presently considered and being evaluated. The first, developed in Russia[5], is a ring switch that has a fraction of 1-nH inductance switching large currents up to ~ 1 MA through multiple conducting channels. It can do that at a low 100-kV voltage. The second is being developed in Sandia[6]. It is a low inductance (~50 pH) high-gain photoconductive semiconductor switch (PCSS) that can switch up to 250-kV, 7-kA currents in a parallel array of six, 2-inch wide GaAs wafers. We believe the photoconducting switch can be further improved to increase the current per unit width by at least one order of magnitude for the same size. The Russian switch has already reached maturity, and the design and hardware are readily available. Those switches make it possible to switch at low, 100-kV voltage and still produce very narrow 50-ns pulses directly from the capacitors without the need of cumbersome and expensive pulse compression.

The fastest capacitors available to date are the compact Maxwell Laboratories S type capacitors model 31165 [7]. They are very small (58x150x274 mm$^2$) and very fast (L=25 nH and C=40 nF). These capacitors could easily be modified to the faster ones (L=25 nH, C= 11 nF) considered in the present point design. Although the capacitor dimensions could be further reduced or special semi-circular geometries could be developed, for the purpose of our point design the dimensions of the presently available off-the-shelf 31165 Maxwell capacitors were assumed.

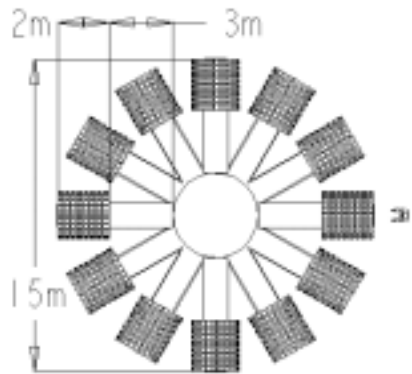

Fig. 2 Cross-sectional view of the new design

The accelerator gaps are magnetically insulated with Metglas.$^{TM}$ A very small Metglas$^{TM}$ cross-sectional area (10 cm$^2$) is needed because of the modest, 2.4x10$^{-3}$ voltsecs of each gap. Therefore, the dimensions of each module are as follows: The overall diameter is 2.2 m and the length 2.4 m (Fig. 2). The cathode electrode is conical starting with a diameter of 1.17 m and terminating at the output end with a 1.08 m diameter cylinder. The anode electrode is a 1.20-m inner diameter cylinder. A 4-m coaxial MITL vacuum transmission line connects each module to the respective conical triplate MITL of each diode ring. The coaxial MITL is longer than the minimum required (2 m) in order to provide easy accessibility and servicing of the accelerator. Throughout the power flow from the voltage adders to the diode rings self-magnetic insulation and impedance matching was rigorously implemented. The power partition ratio of 3:2:1 for the diode rings was retained since the same diode as the one presently operating was assumed to be utilized

One of the advantages of the proposed design is that the diode insulating stack of the central section is eliminated. The only insulator left is the short plastic ring insulating each of the ~60-kV accelerating gaps of the modules. A total of 24 modules are required to provide the 15 MA current to the diode. To maintain the proper power ratio, 4, 8, and 12 modules are connected respectively to the top, middle and bottom conical triplate MITL's. A minimum anode-cathode gap of 1 cm was maintained throughout the entire power flow transport.

Table 1: Comparison of the two designs' performance

| Level | Zsource Ohms | Zload Ohms | Vload MV | I load MA |
|---|---|---|---|---|
| **New Point Design** | | | | |
| Top | 0.91 | 0.99 | 2.5 | 2.5 |
| Middle | 0.45 | 0.38 | 2.2 | 5.7 |
| Bottom | 0.30 | 0.33 | 2.4 | 7.4 |
| **Present Accelerator Performance** | | | | |
| Top | 0.66 | 0.99 | 2.1 | 1.8 |
| Middle | 0.33 | 0.38 | 2.1 | 3.9 |
| Bottom | 0.22 | 0.33 | 2.1 | 6.2 |

Table 1 summarizes the peak pulse power parameter of the new point design and compares it with the present accelerator performance. The design was done analytically and verified numerically utilizing the circuit design code SCREAMER[8].

Figure 3 provides the voltage current and power for the outer-diode ring connected to the bottom MITL.

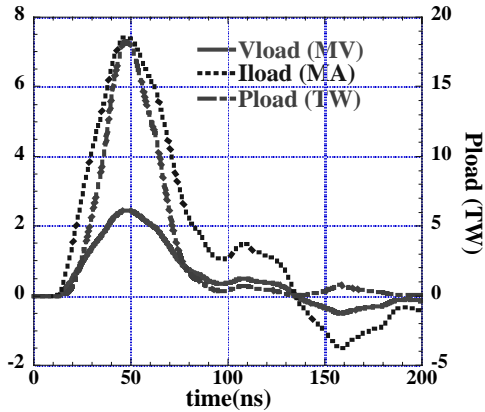

Fig. 3 Outer diode ring (bottom MITL) waveforms.

Similar pulse waveforms are applied to the other two diode rings. Figure 4 is a three dimensional rendition of the entire accelerator.

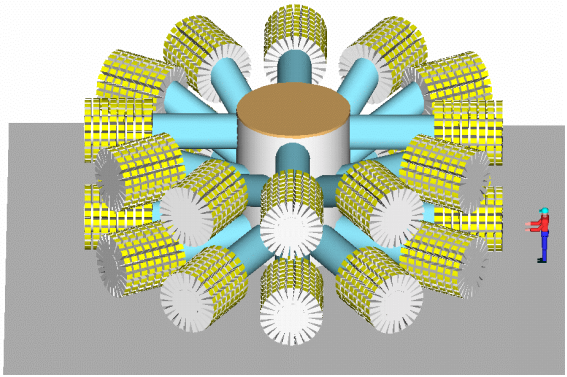

Fig. 4 A 3 D visualization of the new Saturn point design

The same design can deliver to a Z-pinch load a maximum total-current of 15 MA. In this case only 2 conical MITLs will be utilized with each connected to 12 voltage-adder modules. Details of this application and design optimization will be presented in future publication.

## 5 SUMMARY

We have developed a point design for an alternative pulsed power driver for the Saturn accelerator. It has the same or higher power and energy output as the one presently in operation. However, its size is appreciably reduced from the present 30-m diameter to 17 m. The overall height remains the same and of the order of 5 m. The expensive large amounts of oil and deionized water are eliminated together with the requirement for a central water-vacuum interface-insulating stack. A first cut estimate of the cost suggests a relatively modest value of ~$10 M. A cheaper device of ~$5 M could be built if the Metglass$^{TM}$ cores are substituted with air core equivalents. Because of the very short pulse 40-ns FWHM, the dimensions of an air core device will not be appreciably larger than our Metglass$^{TM}$ point design